\documentclass[twocolumn,showkeys,aps,prb,showpacs]{revtex4-1}
\UseRawInputEncoding
\usepackage{graphicx}
\usepackage[CJKbookmarks,dvipdfm,colorlinks,linkcolor=blue,citecolor=blue]{hyperref}

\begin{document}

\title{Janus monolayer ScXY (X$\neq$Y=Cl, Br and I) for piezoelectric and  valleytronic application: a first-principle prediction}

\author{San-Dong Guo$^{1}$, Xiao-Shu Guo$^{1}$, Shuo-Ning Si$^{1}$, Kai Cheng$^{1}$, Ke Wang$^{1}$  and Yee Sin Ang$^{2}$}
\affiliation{$^1$School of Electronic Engineering, Xi'an University of Posts and Telecommunications, Xi'an 710121, China}
\affiliation{$^2$Science, Mathematics and Technology (SMT), Singapore University of Technology and Design (SUTD), 8 Somapah Road, Singapore 487372, Singapore}
\begin{abstract}
Coexistence of  ferromagnetism, piezoelectricity and valley in two-dimensional (2D) materials is crucial to advance multifunctional electronic technologies. Here, Janus ScXY (X$\neq$Y=Cl, Br and I) monolayers are  predicted to be in-plane piezoelectric ferromagnetic (FM) semiconductors with dynamical, mechanical and thermal  stabilities. The predicted  piezoelectric strain coefficients $d_{11}$ and $d_{31}$ (absolute values) are higher than ones of most 2D materials. Moreover, the $d_{31}$ (absolute value) of ScClI reaches up to 1.14 pm/V, which is highly desirable for ultrathin piezoelectric device application.  To obtain spontaneous valley polarization,  charge doping are explored to tune the direction of magnetization of ScXY. By appropriate hole doping, their easy magnetization axis can change from in-plane to out-of-plane,  resulting in spontaneous valley polarization. Taking ScBrI with  0.20  holes per f.u. as a example, under the action of an in-plane electric field, the  hole carriers  of K valley  turn towards one edge of the sample,  which will produce anomalous valley Hall effect (AVHE), and  the hole carriers of $\Gamma$ valley move in a straight line.
These  findings   could pave the way  for designing piezoelectric and valleytronic devices.

\end{abstract}
\keywords{Piezoelectricity, Ferromagnetism, Valley, 2D materials~~~~~~~~~~~~~~~~Email:sandongyuwang@163.com}

\maketitle

\section{Introduction}
In two-dimensional (2D) materials, there may be piezoelectric, magnetic or valley properties, and their coupling may provide a potential platform for multifunctional electronic devices\cite{a1}. The combination of piezoelectricity and ferromagnetic (FM) order, namely 2D piezoelectric ferromagnetism (PFM), has been widely investigated by the first-principle calculations\cite{qt1,q15,q15-0,q15-1,q15-2,q15-3}.
Especially, the PFM with large out-of-plane piezoresponse is highly desirable for ultrathin piezoelectric device application.
Some of PFMs may have ferrovalley (FV) properties with spontaneous spin and valley polarizations\cite{f10}.  In fact, a FV semiconductor with FM ordering must be a PFM due to broken time-reversal and space-inversion symmetries. The piezoelectric properties of some FV semiconductors have been  explored theoretically\cite{f11,f12}.
The anomalous valley Hall effect (AVHE) driven  by piezoelectric effect has been proposed in monolayer $\mathrm{GdCl_2}$\cite{f11}.
Therefore,  searching for  2D PFMs with large vertical piezoelectric response and  FV properties is significative and interesting.

2D Janus materials could be  potential candidates with out-of-plane  piezoelectric response and  FV properties due to broken  out-of-plane symmetry\cite{q16,q16-1}, and the representative Janus  monolayer MoSSe has been experimentally fabricated\cite{e1,e2}.
The Janus strategy has been used to produce large out-of-plane  piezoelectric response\cite{qt1,q15}.  For example Janus monolayer CrSCl, the predicted out-of-plane piezoelectric strain coefficient $d_{31}$ is -1.58 pm/V\cite{q15}. Moreover, for many  three-layer Janus family,  the size of out-of-plane  $d_{31}$ is positively related to electronegativity difference of bottom and top atoms\cite{q15}.

Recently, monolayer  $\mathrm{ScX_2}$ (X = Cl, Br and I) are predicted to be in-plane FM semiconductors, which are dynamically, mechanically and thermally  stable with preserved out-of-plane symmetry\cite{f13}.  These monolayers provide the foundation for building Janus structures to induce out-of-plane piezoelectric response. In this work, we construct Janus  ScXY (X$\neq$Y=Cl, Br and I) monolayers, which are stable in-plane FM semiconductors.
Among the three monolayers, due to largest  electronegativity difference of Cl and I atoms,  ScClI monolayer has the highest $d_{31}$ (absolute value) of 1.14 pm/V, which is higher than ones (less than 1 pm/V) of most 2D materials\cite{q7,q7-2}.  After hole doping, the easy magnetization axis changes from in-plane to out-of-plane,  and  the AVHE can be realized under the action of an in-plane electric field. Taking ScBrI  with 0.20  holes per f.u. as a example,  the carrier doping density is around 1.53$\times$$10^{14}$ $\mathrm{cm^2}$, which can be realized in Hall devices through available gate techniques. Our works provide potential candidate materials for piezoelectric and  valleytronic application.

\section{Computational detail}
 Within density functional theory (DFT)\cite{1}, we perform the first-principle calculations  by using the projector augmented wave (PAW) method as implemented in Vienna ab initio Simulation Package (VASP)\cite{pv1,pv2,pv3}. The generalized gradient approximation  of Perdew, Burke and  Ernzerhof  (GGA-PBE)\cite{pbe} is adopted as  exchange-correlation  functional. The  on-site Coulomb correlation of Sc-3$d$ electrons is considered by PBE+$U$ method within the rotationally invariant approach proposed by Dudarev et al\cite{u}, and  the  $U$$=$2.5 eV\cite{f13} is adopted.
The spin-orbital coupling (SOC) is included to investigate magnetocrystalline anisotropy (MCA) energy and valley properties of ScXY (X$\neq$Y=Cl, Br and I).
The energy cut-off of 500 eV, total energy  convergence criterion of  $10^{-8}$ eV and force
convergence criteria  of less than 0.0001 $\mathrm{eV.{\AA}^{-1}}$ are set to  attain reliable results.
A vacuum space of
more than 16 $\mathrm{{\AA}}$ between slabs along the $z$ direction is added to eliminate the spurious
interactions.
The phonon spectrum  is obtained  by using the  Phonopy code\cite{pv5} with a 5$\times$5$\times$1 supercell.
The elastic  stiffness ($C_{ij}$) are obtained by strain-stress relationship (SSR),  and the piezoelectric stress  tensors  ($e_{ij}$)   are calculated by using density functional perturbation theory (DFPT) method\cite{pv6}.
The Curie temperature is
estimated  by Monte Carlo (MC) simulations with a 40$\times$40  supercell and   $10^7$ loops using
Wolff algorithm, as implemented in Mcsolver code\cite{mc}.
 A 21$\times$21$\times$1 k-point meshes in the first
Brillouin zone (BZ) are used  for electronic structures and $C_{ij}$, and 12$\times$21$\times$1 for FM/antiferromagnetic (AFM) energies and $e_{ij}$.

\begin{figure}
  \includegraphics[width=7cm]{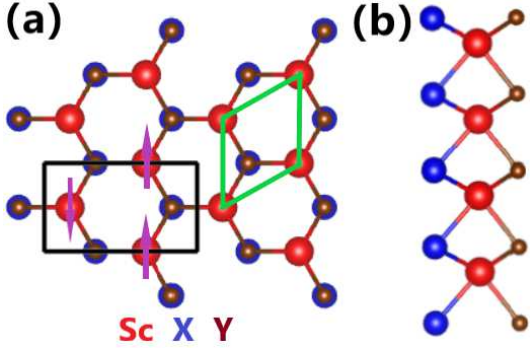}
  \caption{(Color online)For Janus  monolayer ScXY (X$\neq$Y=Cl, Br and I),  the top view (a) and  side view (b) of  crystal structure.  In (a), the primitive (rectangle supercell) cell is
   marked by green (black) lines, and the AFM configuration is shown with arrows as the spin direction of Sc atoms.}\label{t0}
\end{figure}

\begin{figure}
  \includegraphics[width=8cm]{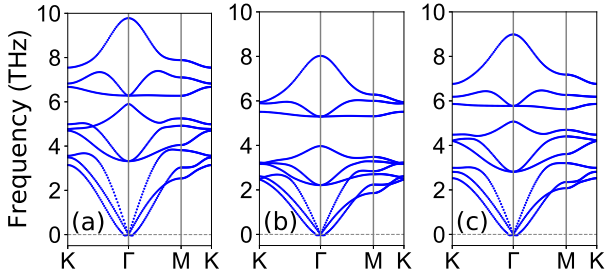}
  \caption{(Color online)The phonon dispersion curves of ScClBr (a), ScBrI (b) and ScClI(c).}\label{t1}
\end{figure}
\begin{table*}
\centering \caption{For ScXY (X$\neq$Y=Cl, Br and I), the lattice constants $a_0$ ($\mathrm{{\AA}}$), the energy difference $\Delta E$ (meV) between AFM and FM ordering with rectangle supercell, the magnetic anisotropy energy MAE (MSA and MCA energies) ($\mathrm{\mu eV}$), the BKT magnetic transition temperature $T_{BKT}$,  the elastic constants $C_{ij}$ ($\mathrm{Nm^{-1}}$), the  piezoelectric coefficients   $e_{ij}$  ($10^{-10}$ C/m ) and  $d_{ij}$ (pm/V). }\label{tab0}
  \begin{tabular*}{0.96\textwidth}{@{\extracolsep{\fill}}ccccccccccc}
  \hline\hline
 Name &$a_0$&$\Delta E$ & MAE (MSA, MCA)& $T_{BKT}$&$C_{11}$&$C_{12}$&$e_{11}$&$e_{31}$&$d_{11}$&$d_{31}$\\\hline\hline
$\mathrm{ScClBr}$ &   3.649  &76      & -53 (-3, -50)  &147  &48.19    &15.72    &-2.43 &-0.28   &-7.49  &-0.44                           \\\hline
$\mathrm{ScBrI}$ &   3.881  &87     & -228 (-3, -225)  &168  &41.58   &12.92    &-2.10  &-0.43  &-7.33  &-0.79                       \\\hline
$\mathrm{ScClI}$ &    3.806   &77    & -198 (-3, -195) &149  &46.48    &15.13    &-2.32   &-0.70   &-7.39   &-1.14
\\\hline\hline
\end{tabular*}
\end{table*}

\section{Main calculated results}
For Janus  monolayer ScXY (X$\neq$Y=Cl, Br and I), the crystal structures are shown in \autoref{t0}.  The Sc atoms are surrounded by three X and three Y atoms, and  form a distorted
octahedral structure due to  inequivalent Sc-X and Sc-Y bonding lengths caused by different atomic size and electronegativity of X and Y
atoms. This also leads to lower space group of $P3m1$ (No.156) with respect to monolayer $\mathrm{ScX_2}$ (X=Cl, Br and I)\cite{f13}. Therefore, both central inversion symmetry and horizontal mirror symmetry of ScXY are broken, which can induce both in-plane and out-of-plane piezoelectric response.
Their energy differences between AFM and FM configurations ($\Delta E$=$E_{AFM}$-$E_{FM}$) are listed in \autoref{tab0}, and the positive values  confirm that they all are FM ground state.   The optimized lattice constant $a$ with FM ordering are also summarized in \autoref{tab0}.

The magnetocrystalline direction determines  type of magnetic phase transition of 2D hexagonal symmetric system with a typical triangle lattice structure, which can be determined by calculating  magnetic anisotropy energy (MAE),  including MCA energy ($E_{MCA}$)  and  magnetic shape anisotropy (MSA) energy ($E_{MSA}$).
The MCA depends on SOC, and the $E_{MCA}$ can be calculated by $E_{MCA}=E^{||}_{SOC}-E^{\perp}_{SOC}$, where $||$ and $\perp$  mean that spins lie in-plane and out-of-plane.
 The MSA is produced by the anisotropic dipole-dipole (D-D) interaction\cite{a1-7,a7-1}:
 \begin{equation}\label{d-d}
E_{D-D}=\frac{1}{2}\frac{\mu_0}{4\pi}\sum_{i\neq j}\frac{1}{r_{ij}^3}[\vec{M_i}\cdot\vec{M_j}-\frac{3}{r_{ij}^2}(\vec{M_i}\cdot\vec{r_{ij}})(\vec{M_j}\cdot\vec{r_{ij}})]
 \end{equation}
where  the $\vec{M_i}$   represent the local magnetic moments of Sc atoms, and   the sites $i$ and $j$ are connected by vectors $\vec{r_{ij}}$.
For a collinear FM monolayer,  the $E_{MSA}$  ($E_{D-D}^{||}-E_{D-D}^{\perp}$)can be written as :
\begin{equation}\label{d-d-3}
E_{MSA}=-\frac{3}{2}\frac{\mu_0M^2}{4\pi}\sum_{i\neq j}\frac{1}{r_{ij}^3}\cos^2\theta_{ij}
 \end{equation}
 where $\theta_{ij}$ is the angle between the $\vec{M}$ and $\vec{r_{ij}}$. It is clearly seen that  the $E_{MSA}$ depends the crystal structure and local magnetic moment of magnetic atoms, which  tends to make spins be in-plane.

The calculated MAE, MSA and MCA energies of ScXY (X$\neq$Y=Cl, Br and I) are listed in \autoref{tab0}, and the negative MAE indicate that the three monolayers  have an in-plane magnetic anisotropy. It is found that their MSA energies are about -3 $\mathrm{\mu eV}$ with  magnetic moment of Sc atom about 0.59 $\mu_B$, which is very smaller than MCA energy. Therefore, we will ignore MSA energy  in later calculations of MAE.
For in-plane case,  there is no energetic barrier to the rotation of magnetization in the $xy$ plane, and
 a Berezinskii-Kosterlitz-Thouless (BKT) magnetic transition will be produced with a quasi-long-range
phase\cite{re5,re5-1}.   Therefore, the ScXY (X$\neq$Y=Cl, Br and I) are 2D $XY$ magnets. The monte Carlo simulations have shown that BKT magnetic transition takes place at a critical temperature $T_{BKT}=1.335\frac{J}{K_B}$\cite{re6,re7}, where $J$ and $K_B$ are the nearest-neighboring exchange parameter and  Boltzmann
constant.
The $J$ can be  determined from $\Delta E$=$E_{AFM}$-$E_{FM}$, and the  FM and AFM energies with rectangle supercell  can be
obtained:
\begin{equation}\label{pe0-1-2}
E_{FM}=E_0-6JS^2
 \end{equation}
  \begin{equation}\label{pe0-1-3}
E_{AFM}=E_0+2JS^2
 \end{equation}
where $E_0$ means the total energy of system without magnetic coupling.
The  corresponding $J$ can be expressed:
  \begin{equation}\label{pe0-tc}
J=\frac{E_{AFM}-E_{FM}}{8S^2}
 \end{equation}
The predicted $T_{BKT}$ of  ScXY (X$\neq$Y=Cl, Br and I) ($S$=1) are summarized in \autoref{tab0}, which are higher than those of  experimental discovered 2D ferromagnetic
materials $\mathrm{CrI_3}$ (45 K) and $\mathrm{Cr_2Ge_2Te_6}$ (30 K)\cite{f3-2,f3-3}.

The phonon dispersions of  ScXY (X$\neq$Y=Cl, Br and I) are calculated to  verify their dynamical stabilities,  as shown in  \autoref{t1}.   All the phonon frequencies are non-negative in the whole BZ,  implying   that three monolayers are dynamically stable.
To further confirm their thermal stabilities, the  ab-initio molecular dynamics (AIMD) simulations are performed
 with a 4$\times$4$\times$1 supercell and a time step
of 1 fs at 300 K for 8 ps.  According to FIG.1 of electronic supplementary information (ESI),  the
energy of ScXY (X$\neq$Y=Cl, Br and I) fluctuates  within a small range during the whole simulation time. Moreover, their snapshots  show no structural transitions at the end of the AIMD simulations, manifesting their thermal stabilities.
The linear elastic constants of ScXY (X$\neq$Y=Cl, Br and I) are calculated to determine their mechanical stabilities.
Due to  $P3m1$ space group, only
two independent elastic constants ($C_{11}$ and $C_{12}$) can be observed, which are shown in \autoref{tab0}.
These $C_{11}$ and $C_{12}$ meet Born-Huang
criteria of  mechanical stability  ($C_{11}>0$ and  $C_{11}-C_{12}>0$)\cite{ela},  thereby verifying their mechanical
stabilities.

\begin{figure}
     \includegraphics[width=8cm]{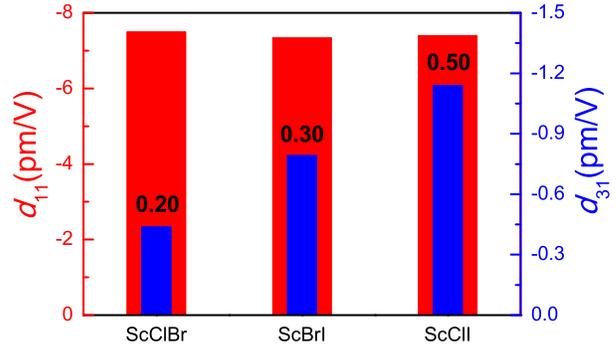}
  \caption{(Color online)For ScXY (X$\neq$Y=Cl, Br and I), the piezoelectric strain coefficients $d_{11}$ and $d_{31}$, and the electronegativity difference of X and Y atoms is shown in the bar chart.}\label{t3}
\end{figure}

Due to  unique Janus
structure of ScXY (X$\neq$Y=Cl, Br and I),  the broken central inversion symmetry and horizontal mirror symmetry can induce both in-plane and  out-of-plane piezoelectric response.  By using  Voigt notation, due to $P3m1$ space group, the 2D elastic tensor,  piezoelectric stress and  strain  tensors  can be reduced into\cite{q7,q7-2}:
\begin{equation}\label{pe1-4}
   C=\left(
    \begin{array}{ccc}
      C_{11} & C_{12} & 0 \\
     C_{12} & C_{11} &0 \\
      0 & 0 & (C_{11}-C_{12})/2 \\
    \end{array}
  \right)
\end{equation}

 \begin{equation}\label{pe1-1}
 e=\left(
    \begin{array}{ccc}
      e_{11} & -e_{11} & 0 \\
     0 & 0 & -e_{11} \\
      e_{31} & e_{31} & 0 \\
    \end{array}
  \right)
    \end{equation}

  \begin{equation}\label{pe1-2}
  d= \left(
    \begin{array}{ccc}
      d_{11} & -d_{11} & 0 \\
      0 & 0 & -2d_{11} \\
      d_{31} & d_{31} &0 \\
    \end{array}
  \right)
\end{equation}
    With an imposed  uniaxial in-plane strain,  $e_{11}$/$d_{11}$$\neq$0 and $e_{31}$/$d_{31}$$\neq$0. However,  $e_{11}$/$d_{11}$=0, but $e_{31}$/$d_{31}$$\neq$0, when  a biaxial in-plane strain is applied. Here, the two independent $d_{11}$ and $d_{31}$ can be derived by $e_{ik}=d_{ij}C_{jk}$:
\begin{equation}\label{pe2}
    d_{11}=\frac{e_{11}}{C_{11}-C_{12}}~~~and~~~d_{31}=\frac{e_{31}}{C_{11}+C_{12}}
\end{equation}
\begin{figure}
     \includegraphics[width=8cm]{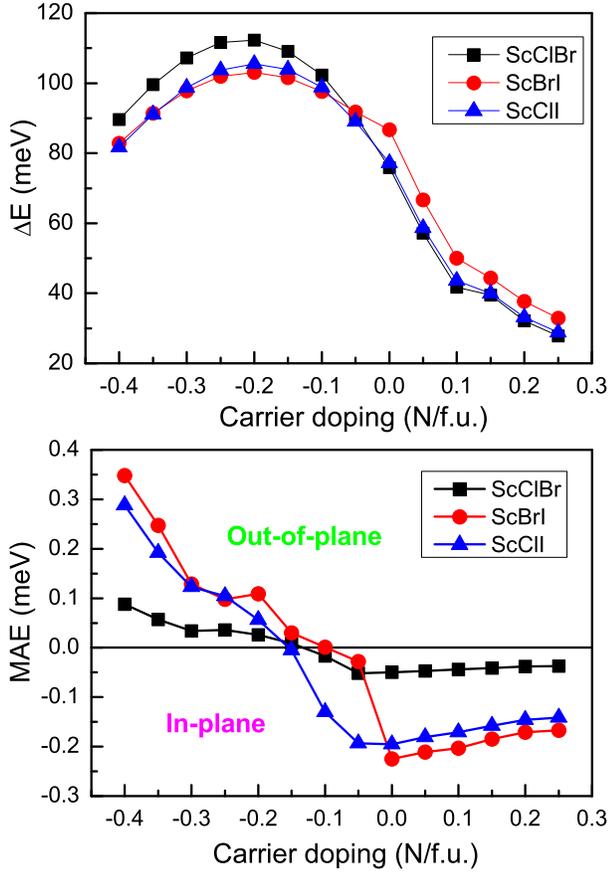}
  \caption{(Color online)For ScXY (X$\neq$Y=Cl, Br and I), the energy differences $\Delta E$ between  AFM and FM ordering and MAE  as a function of carrier doping concentration.}\label{t4}
\end{figure}

A piezoelectric material should be a semiconductor to  prevent leakage current. The energy band structures of ScXY (X$\neq$Y=Cl, Br and I) with intrinsic in-plane magnetization are calculated, as shown in FIG.2 of ESI. It is clearly seen that they all are indirect bandgap semiconductors.
The conduction band bottom (CBM) of three monolayers all are at M point, and the valence band maximum (VBM) locates at $\Gamma$ point except for ScBrI at K/-K point.
The  orthorhombic supercell (see  \autoref{t0}) is used to calculate the  $e_{11}$/$e_{31}$ of ScXY (X$\neq$Y=Cl, Br and I).
The elastic constants  ($C_{11}$, $C_{12}$,  $C_{11}$-$C_{12}$ and $C_{11}$+$C_{12}$) and  piezoelectric  stress  coefficients  ($e_{11}$ and $e_{31}$) along  the ionic  and electronic contributions of ScXY (X$\neq$Y=Cl, Br and I) are shown in FIG.3 of ESI.
For $e_{11}$ of three monolayers, the electronic and ionic parts  have  superposed contributions, and the ionic part is larger than electronic one. However,   the electronic and ionic contributions of $e_{31}$ have  opposite signs, and   the electronic part dominates the  piezoelectricity.

And then, their  $d_{11}$/$d_{31}$ of ScXY can be calculated from \autoref{pe2}, which are plotted in \autoref{t3} along with the electronegativity difference between X and Y atoms. It is found that the $d_{11}$ of three monolayers are almost the same (absolute value for about 7.40 pm/V), which is higher than ones of many 2D materials\cite{q7,q7-2}.  A large out-of-plane piezoelectric response (for example high $d_{31}$) is
highly desired to be compatible with the
nowadays bottom/top gate technologies to meet the needs of practical application,.
For many 2D Janus families, the size of out-of-plane response $d_{31}$ is positively related to electronegativity difference of bottom and top atoms\cite{q15}. It is clearly seen that ScXY family is also accord with this law. From ScClBr to ScBrI to ScClI, the electronegativity difference of X and Y increases, and then the $d_{31}$ enhances. For ScClI, the $d_{31}$ (absolute value) reaches up to 1.14 pm/V, which is higher ones of most 2D materials ($<$1 pm/V), such as  oxygen functionalized MXenes (0.40-0.78 pm/V)\cite{q9},  Janus TMD monolayers (0.03 pm/V)\cite{q7},
functionalized h-BN (0.13 pm/V)\cite{o1}, kalium decorated graphene (0.3
pm/V)\cite{o2}, Janus group-III materials (0.46 pm/V)\cite{q7-6-1}, Janus BiTeI/SbTeI  monolayer (0.37-0.66 pm/V)\cite{o3}, $\alpha$-$\mathrm{In_2Se_3}$
(0.415 pm/V)\cite{o4} and MoSO (0.7 pm/V)\cite{re-11}.
The large out-of-plane piezoelectric response also has been predicted in some 2D materials, such as NiClI (1.89 pm/V)\cite{qt1}, TePtS/TePtSe (2.4-2.9 pm/V)\cite{re-6}, CrSCl (1.58 pm/V)\cite{q15} and $\mathrm{CrBr_{1.5}I_{1.5}}$ (1.138 pm/V)\cite{q15-1}. However,   the ScClI possesses  FM ordering with large in-plane piezoelectric response with respect to these 2D materials.

\begin{figure}
     \includegraphics[width=8cm]{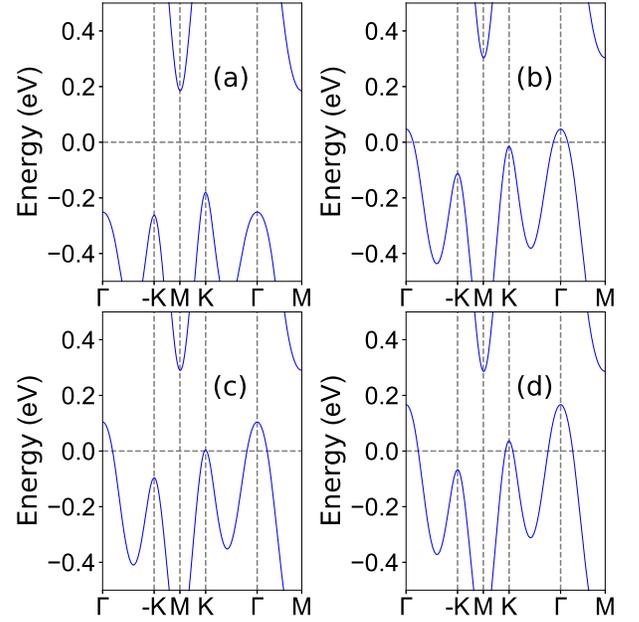}
  \caption{(Color online)For ScBrI monolayer, the energy band structures with doping 0.00 (a), 0.10 (b), 0.15 (c) and 0.20 (d) holes per f.u..}\label{t5}
\end{figure}

\begin{figure*}
     \includegraphics[width=12cm]{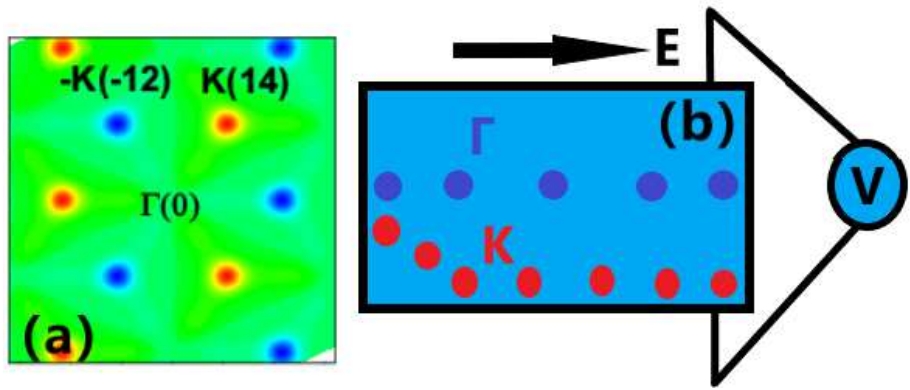}
  \caption{(Color online)(a): For undoped case, the Berry curvature of ScBrI with assumed out-of-plane direction in a contour map in the 2D BZ. The Berry curvature mainly occurs around -K and K valleys with opposite
signs and unequal magnitudes, and  Berry curvature around $\Gamma$ valley is almost zero. (b): For 0.20  holes per f.u., under an in-plane longitudinal electric field $E$,
    the  hole carriers  of K valley  turn towards one edge of the sample, and  the hole carriers of $\Gamma$ valley move in a straight line. }\label{t6}
\end{figure*}
\begin{figure}
     \includegraphics[width=8cm]{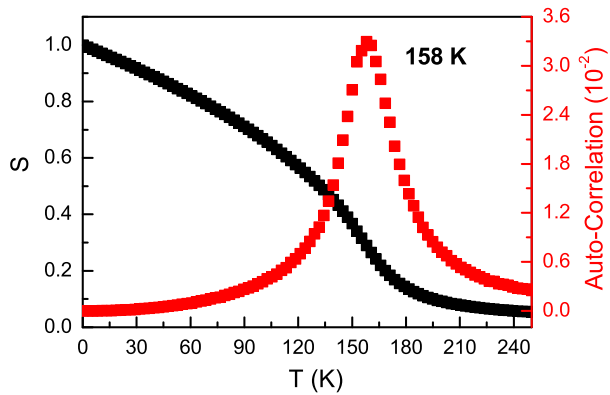}
  \caption{(Color online)For  monolayer ScBrI with 0.20  holes per f.u., the  magnetic moment (S) and auto-correlation  as a function of temperature. }\label{tc-1}
\end{figure}

According to the Sc-$d$ band structure projection of ScXY (FIG.2 of ESI), the -K and K valleys of valence band are near the Fermi level with $d_{xy}$/$d_{x^2-y^2}$ characters.  This means that the spontaneous  valley polarization will be produced, when the magnetization direction
 is tuned to out-of-plane case by external field, such as magnetic field, electric field and carrier doping\cite{f13,f16,f17}.  Here, the carrier doping  with the concentration of -0.4$\sim$0.25 carriers per f.u. is used to tune
magnetization direction of ScXY, and the negative/positive values mean the hole/electron doping.
For ScXY (X$\neq$Y=Cl, Br and I), the energy differences $\Delta E$ between  AFM and FM orderings and MAE  as a function of carrier doping concentration are plotted in \autoref{t4}. It is found that the FM ordering is always ground state in considered doping range for three monolayers, and
 the FM interaction can be enhanced in  a certain hole doping range.
In considered doping range, the MAE of SxXY
is always negative for the electron doping, indicating that their
magnetization  direction is always in-plane. Fortunately, the easy axis becomes out-of-plane, when
 the hole doping concentration is beyond about 0.15/0.10/0.15 holes per f.u. for ScClBr/ScBrI/ScClI.

 The energy band structures of monolayer ScBrI with representative hole doping are plotted in \autoref{t5}, and those of ScClBr and ScClI are shown in FIG.4 and FIG.5 of ESI.  For undoped ScBrI monolayer, the energy at K point is higher than one at $\Gamma$ point. The hole doping make the energy at $\Gamma$ point become higher than one at K point. When the hole doping concentration is over 0.15 holes per f.u, the K valley is also doped.
 For undoped case, the Berry curvature distribution of ScBrI with assumed out-of-plane direction is plotted in \autoref{t6}. The Berry curvature mainly occurs around -K and K valleys with opposite
signs and unequal magnitudes, and  Berry curvature around $\Gamma$ valley is almost zero.
Under an in-plane longitudinal electric field $E$,
the  carriers with the nonzero Berry curvature $\Omega (k)$ can obtain the general group velocity $v_{\parallel}$ and
anomalous transverse velocity $v_{\bot}$\cite{xd,qqq}:
\begin{equation}\label{d-d-1}
v=v_{\parallel}+v_{\bot}=\frac{1}{\hbar}\nabla_k\varepsilon(k)-\frac{e}{\hbar}E\times\Omega(k)
 \end{equation}
where $v_{\parallel}$  is along the electric field direction, and  $v_{\bot}$  is perpendicular to the electric field and out-of-plane directions.
For ScBrI with 0.20  holes per f.u., under an in-plane longitudinal electric field $E$, the  hole carriers  of K valley  turn towards one edge of the sample due to anomalous transverse velocity $v_{\bot}$,  which will induce a
charge Hall current with a voltage (namely AVHE), and  the hole carriers of $\Gamma$ valley move in a straight line (see \autoref{t6}).
 The corresponding carrier doping density is around 1.53$\times$$10^{14}$ $\mathrm{cm^2}$, which can be experimentally  realized by available gate techniques.

The Curie temperature $T_C$ of ScBrI with 0.20  holes per f.u. is estimated with  the spin Hamiltonian under
the Heisenberg model:
\begin{equation}\label{pe0-1-1}
H=-J\sum_{i,j}S_i\cdot S_j-A\sum_i(S_i^z)^2
 \end{equation}
where   $J$, $S$ and $A$ are  the nearest exchange parameter,  spin quantum number and  MAE, respectively.
 The $J$ can be calculated by \autoref{pe0-tc}, and the corresponding value is 12.88 meV. The   magnetic moment and auto-correlation  as a function of  temperature  are plotted in \autoref{tc-1}, and the  $T_C$ is predicted to be  about 158 K,  which is higher than those of 2D ferromagnetic
materials $\mathrm{CrI_3}$ (45 K) and $\mathrm{Cr_2Ge_2Te_6}$ (30 K)\cite{f3-2,f3-3}.

\section{Conclusion}
In conclusion, the  piezoelectric and valley
properties of Janus ScXY (X$\neq$Y=Cl, Br and I)  are systematically studied by first-principles calculations.
Calculated results show that these monolayers are in-plane FM semiconductors with excellent stability.
The findings indicate that the  piezoelectric and FM properties  can be combined in monolayer ScXY (X$\neq$Y=Cl, Br and I), and both in-plane and out-of-plane piezoelectric coefficients are higher than those of most known 2D materials.
 Under appropriate holes doping, the easy magnetization axis of these monolayers turns to out-of-plane from in-plane, and the AVHE can be realized under the action of an in-plane electric field.  Our works  provide  a development guide for  searching 2D multifunctional materials.
~~~~\\
~~~~\\
\textbf{SUPPLEMENTARY MATERIAL}
\\
See the supplementary material for AIMD results, projected energy band structures and  elastic/piezoelectric stress  coefficients of ScXY (X$\neq$Y=Cl, Br and I), and the energy band structures of ScClBr and ScClI with hole doping.

~~~~\\
~~~~\\
\textbf{Conflicts of interest}
\\
There are no conflicts to declare.

\begin{acknowledgments}
This work was supported by Natural Science Basis Research Plan
in Shaanxi Province of China (No. 2021JM-456). We are grateful to
Shanxi Supercomputing Center of China, and the calculations were
performed on TianHe-2.
\end{acknowledgments}

\end{document}